\documentclass[a4paper]{spie}  

\usepackage{amsmath,amsfonts,amssymb}
\usepackage{graphicx}
\usepackage[colorlinks=true, allcolors=blue]{hyperref}
\usepackage{siunitx}
\usepackage{float} 
\usepackage{placeins}
\usepackage{acro}
\usepackage{subcaption}
\ExplSyntaxOn
\NewDocumentCommand{\ccite}{m}
 {
  \seq_set_split:Nnn \l_tmpa_seq { , } { #1 }
  \int_compare:nNnTF { \seq_count:N \l_tmpa_seq } > { 1 }
    {[\citenum{#1}] }
    {[\citenum{#1}] }
 }
\ExplSyntaxOff

\title{Characterisation of the NewAthena WFI's DEPFET Flight Production's Operational Parameters}

\author[a]{Léonie~Sommer}
\author[a]{Johannes~Müller-Seidlitz}
\author[a]{Valentin~Emberger}
\author[a]{Robert~Andritschke}
\author[a]{Astrid~Mayr}
\author[a]{Annika~Behrens}
\author[a]{Günter~Hauser}
\author[b]{Peter~Lechner}
\author[b]{Christian~Sandow}
\author[a]{Anna~Schweingruber}
\author[a]{Jonas~P.~Reiffers}
\author[a]{Elif~Kutdemir~Öncü}

\affil[a]{Max Planck Institute for Extraterrestrial Physics, Gießenbachstraße 1, 85748 Garching bei München, Germany}
\affil[b]{Semiconductor Laboratory of the Max-Planck-Society, Isarauenweg 1, 85748 Garching bei München, Germany}

\authorinfo{Corresponding author: \\L. Sommer: E-mail: lsommer@mpe.mpg.de, Telephone: +49 (0)89 30 000 3052}

\DeclareAcronym{ASIC}{
    short = ASIC, 
    long = application-specific integrated circuit
}
\DeclareAcronym{ADC}{
    short = ADC, 
    long = analog digital converter
}
\DeclareAcronym{WFI}{
    short = WFI,
    long = Wide Field Imager
}
\DeclareAcronym{Athena}{
    short = NewAthena,
    long = New Advanced Telescope for High-ENergy Astrophysics
}
\DeclareAcronym{LD}{
    short = LD,
    long = Large Detector
}
\DeclareAcronym{LDA}{
    short = LDA,
    long = Large Detector Array
}
\DeclareAcronym{FD}{
    short = FD,
    long = Fast Detector
}
\DeclareAcronym{DEPFET}{
    short = DEPFET,
    long = Depleted P-channel Field Effect Transistor
}
\DeclareAcronym{FWHM}{
    short = FWHM,
    long = full width at half maximum
}
\DeclareAcronym{OS}{
    short = OS,
    long = outer substrate
}
\DeclareAcronym{PCB}{
    short = PCB,
    long = printed circuit board
}
\DeclareAcronym{PCM}{
    short = PCM,
    long = power conditioning module
}
\DeclareAcronym{SPO}{
    short = SPO,
    long = silicon pore optics
}
 
\begin{document} 
\maketitle

\begin{abstract}
NewAthena’s Wide Field Imager (WFI) uses detectors made up from Depleted P-Channel Field Effect Transistor (DEPFET) pixels operated in rolling shutter mode. The Large Detector Array (LDA) contains a 2 × 2 array of 512 × 512 pixels Large Detectors (LDs) allowing for a field of view of 40’ × 40’ with a frame time of \SI{2}{\milli \second} while the 64 × 64 pixels Fast Detector (FD) can observe very bright X-ray sources due to a faster frame time of 80 µs. 
Prototype sensors (64 × 64 pixels) were used to analyse the sensor's operational range and to optimise the ASIC and DEPFET parameters i.e. current and voltage settings, as well as the DEPFET read-out timing parameters, resulting in an improved energy resolution, reduced noise and otherwise improved sensor characteristics.
\end{abstract}


\section{Introduction}
\label{sec:introduction}

The \acf{Athena} is set to be ESA's next large X-ray mission featuring a mirror using \ac{SPO} with a focal length of \SI{12}{\meter} achieving a wide field of view, large effective area, and an angular resolution of 9 arc-seconds half-energy width.\ccite{SPOpaper}

The science goals of the \ac{WFI}, which is one of two instruments for \ac{Athena}, are to gain insights into the formation of large scale structures of ordinary matter in the universe, e.g. galaxy clusters as well as studying the formation and growth of black holes, accretion effects and the influence they have on galaxy structure.\ccite{nandra2013hotenergeticuniversewhite}

The \acf{WFI} is designed for imaging and spectroscopy of bright X-ray sources up to and above 1 Crab using two dedicated sensors, the \acf{LDA} and \acf{FD}. The \ac{LDA} is made up of 4 \acfp{LD}, each containing $512 \times 512$ \acf{DEPFET} pixels, while the \ac{FD} has $64 \times 64$ \ac{DEPFET} pixels. 
Both, the \acp{LD} and the \ac{FD}, share the same types of \acp{ASIC}: The Switcher-A is used for row activation as the sensors are operated in rolling shutter mode and the VERITAS is used for column read-out. To achieve the faster timing the \ac{FD} has two VERITAS \acp{ASIC} so that only 32 rows are read out with one VERITAS (split-frame readout) as opposed to 64 rows being read out by one VERITAS \ac{ASIC} for prototype sensors of the same size \ccite{Veritas1, VeritasAnna}.
The \ac{LD} and \ac{FD} sensors have a thickness of $\SI{450}{\micro \meter}$ and a pixel size of $\SI{130}{\micro \meter} \times \SI{130}{\micro \meter}$.\ccite{nandratobepublished, SpieValeria}

The spectral resolution for $\le \SI{7}{\kilo \electronvolt}$ is expected to be $\le \SI{170}{ \electronvolt}$ FWHM (full width at half maximum) with the detector working with high quantum efficiency in the range of \SI{0.2}{\kilo \electronvolt} to \SI{15}{\kilo \electronvolt}\ccite{SpieAstrid}. The sensors are back illuminated and the back side of each sensor is coated with \SI{20}{\nano \meter} of SiO$_2$, \SI{30}{\nano \meter} of Si$_3$N$_4$ and \SI{86.5}{\nano \meter} of Aluminium to act as optical and ultra violet blocking filter for lower energy photons.\ccite{Müller_Seidlitz_2024, WFIinternal}

The main voltages and currents applied to the sensor and \acp{ASIC} and the parameters of the read-out sampling are varied. The results of these measurements are analysed to investigate the operational range of the detector and identify the parameter settings for optimised sensor performance.
\section{Prototype DEPFET sensor and testing setup}
\label{sec:sensor&setup}
\FloatBarrier

Prototype sensors with the identical pixel design as the \ac{LD} and \ac{FD} sensors have been developed for testing and are produced on the same wafers. They have $64 \times 64$ pixels and need one Switcher-A and one VERITAS \ac{ASIC} to be controlled and read out. 

\begin{figure}[htb!]
    \centering
    \includegraphics[width=0.6\linewidth]{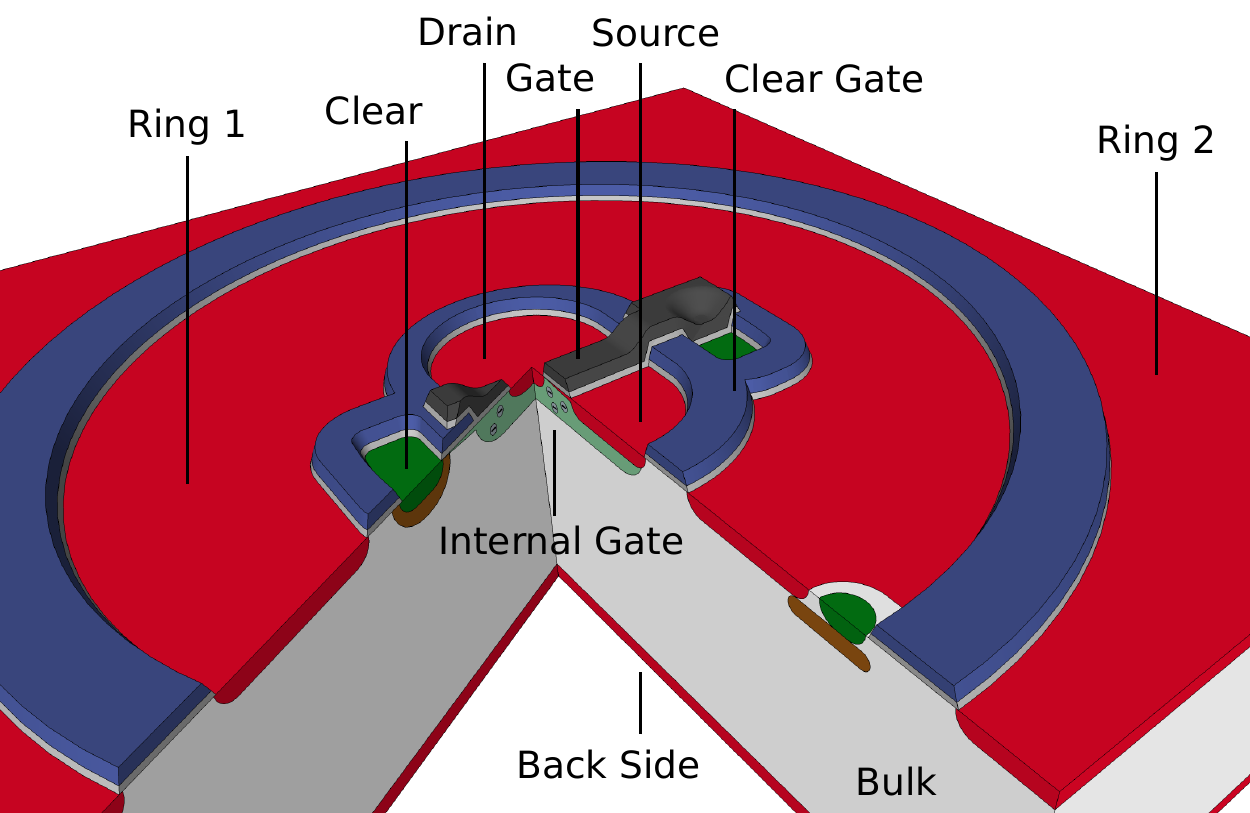}
    \caption{3D layout of one WFI pixel\ccite{WFIinternal}.}
    \label{fig:athenapixel}
\end{figure}

Figure \ref{fig:athenapixel} shows a 3D layout of one \ac{WFI} pixel. 
When an X-ray photon hits the sensitive detector volume, electron-hole pairs are created, which are separated by a voltage applied over the sensor bulk. The holes are drained by the \textit{back contact}, meanwhile the electrons are trapped in a potential well below the transistor \textit{gate} called the \textit{internal gate} and generate mirror charges in the transistor channel.
The transistor \textit{gate}, \textit{source} and \textit{drain} can be seen in the centre of figure \ref{fig:athenapixel}.
The \textit{internal gate} is pictured below the \textit{gate}. Next to the \textit{gate} the \textit{clear} and \textit{clear gate} are located which, when switched on, drain electrons from the \textit{internal gate}.
This entire structure is surrounded by \textit{ring 1} and \textit{ring 2} which focus electrons towards the \ac{DEPFET}.
On the back side the \textit{back contact} for bulk depletion is surrounded by a guard ring called the \textit{back contact inner guard ring}. This ring surrounds the square containing the $64 \times 64$ \ac{DEPFET} pixels. On the front side a similar guard ring called \textit{\acf{OS}} is located. Both guard rings prevent unwanted currents from flowing in and out of the sensitive area of the sensor. Outside of the \textit{\ac{OS}}, a temperature diode is located so that the sensor temperature can be measured.

\begin{figure}[H]
    \centering
    \includegraphics[width=0.525\linewidth]{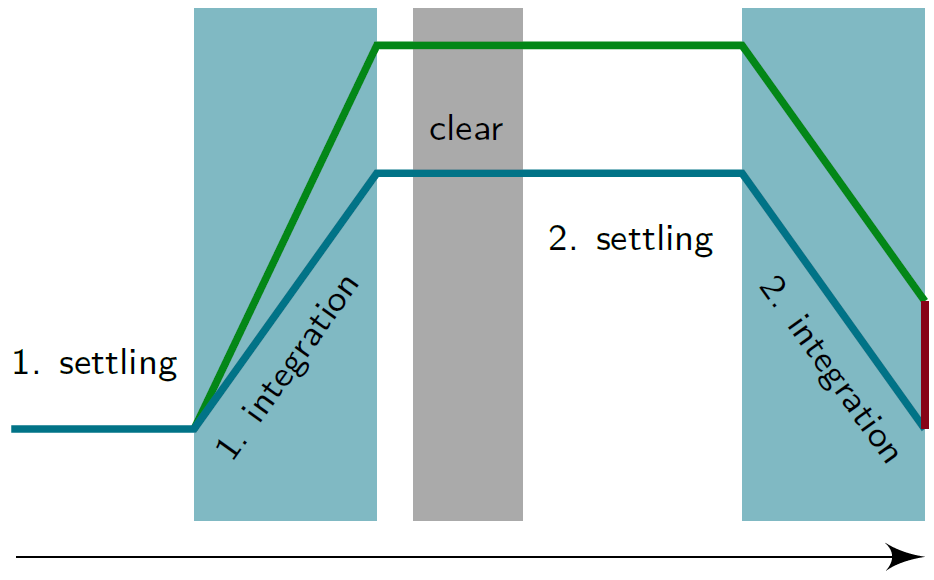}
    \caption{Illustration of the DEPFET readout scheme with (green) and without (blue) collected electrons at the \textit{internal gate} \ccite{JohannesPhD}.}
    \label{fig:Johannestiming}
\end{figure}

Correlated double sampling is used to read out the DEPFETs. When a row is activated, the VERITAS ASIC needs time to settle into a stable state, which is designated as \textit{settling 1}. The first integration allows to determine the charge carriers that form the photon signal and the baseline signal. Then the \textit{clear} depletes the \textit{internal gate} from charge carriers. A second settling (\textit{settling 2}) and integration with empty \textit{internal gate} are then performed to determine the baseline signal. The VERITAS ASIC subtracts the baseline signal from the photon and baseline signal measured first, resulting in only the signal being further processed. Figure \ref{fig:Johannestiming} shows an illustration of this process with electrons collected at the \textit{internal gate} (green line) and without (blue line).\ccite{JohannesPhD}

\begin{figure}[htb!]
\centering
\begin{minipage}[t]{0.49\textwidth}
\centering
\includegraphics[height=5cm]{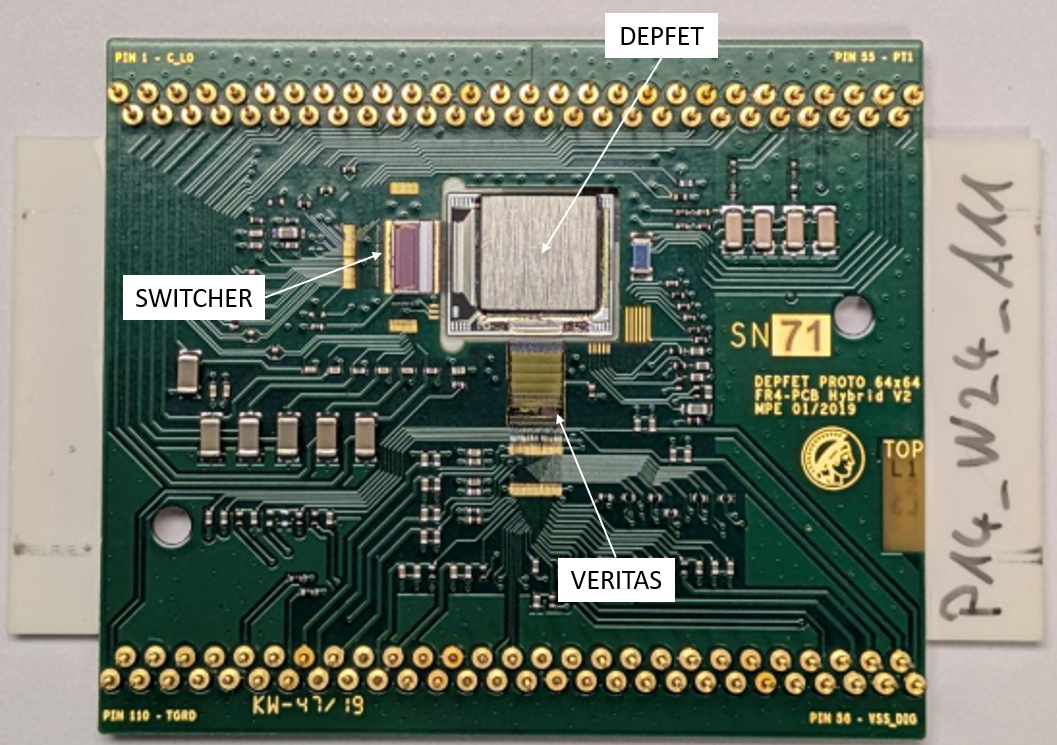}
\label{}
\subcaption{front side}
\end{minipage}
\hfill
\begin{minipage}[t]{0.49\textwidth}
\centering
\includegraphics[angle=90, height=5cm]{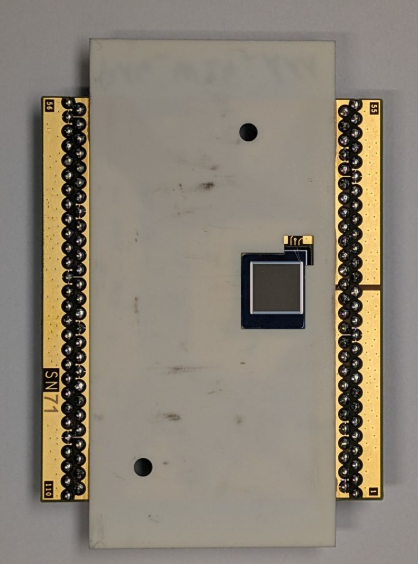}
\subcaption{back side}
\end{minipage}
\caption{Front and back side of the prototype \ac{DEPFET} module used for the characterisation.}
\label{fig:P14_W24_A11_VH}
\end{figure}

For testing, a detector module was used that contained a prototype sensor connected to a Switcher-A and VERITAS \ac{ASIC} on a \ac{PCB}. A ceramic cooling plate is attached to the back side of the PCB to ensure that the sensor is adequately cooled. The module used for testing is shown in figure \ref{fig:P14_W24_A11_VH}.
The detector module is mounted onto an adapter board inside a vacuum chamber, which is equipped with a backing vacuum pump, a turbomolecular pump, and a Stirling cooler ensuring a controlled environment. The back side of the vacuum chamber hosts feed-throughs for the power supply, read-out, and steering. An $^{55}$Fe source is used as X-ray source for measurements.

By default, the DEPFET \textit{source} voltage is set to \SI{5.0}{\volt} and the \textit{drain} voltage to \SI{1.0}{\volt}. The \textit{gate on} voltage is set so that a current of $\sim \SI{6.4}{\milli \ampere}$ flows through the pixel. This usually results in a value of $V_{\text{Gate on}} \sim \SI{3}{\volt} $. The \textit{gate off} voltage is set to $V_{\text{Gate off}} = V_{\text{Gate on}} + \SI{7}{\volt} $. The voltage of the VERITAS current generator $V_{\text{VSSS}}$ is set to $\sim \SI{12}{\volt}$ such that the magnitude of the current $I_{\text{VSSS}}$ is equal to the current running through the pixel.

Generally, the sensor is cooled to approximately \SI{-60}{\celsius}. Unless stated otherwise, the sensor was operated using a row time of either \SI{3.9}{\micro \second} or \SI{5.0}{\micro \second} and sometimes an exposure time of usually \SI{1267.5}{\micro \second} was introduced to register more events with lower frame numbers.

For data analysis the ROOT based Offline and Online Analysis (ROAn) framework \ccite{ROAn} is used. Parameters derived using ROAn do not come with error margins, as the computation is already quite extensive. The noise is often given as average of all ($64 \times 64$) pixels, therefore the error is small. The energy resolution values (\ac{FWHM}) are obtained from fits over the calibrated spectra, which introduces additional error sources.

\FloatBarrier
\section{Current and Voltage variations}
\FloatBarrier

The voltages and currents investigated are shown in table \ref{tab:power-up sequence}. The intervals over which each parameter was varied is also given. As the step size varied for each measurement and additional measurements were taken in areas of interest the exact information on the variation is described in detail in the corresponding section. The default values before and after characterisation are also indicated.

\begin{table}[H]
    \centering
    \caption{Intervals, in which the voltages and currents investigated, were varied. The old and new default values are also shown.}
    \begin{tabular}{|l|c|c|c|}
    \hline
    Parameter  & Variation & Old default & New default    \\
    \hline
    Back contact inner guard ring & $[\SI{-90}{\volt}, \, \SI{-100}{\volt}],  \SI{-105}{\volt} , \SI{-109}{\volt}$  &\SI{-90.0}{\volt} &\SI{-90.0}{\volt} \\
    Drain-source voltage & $[\SI{-3.0}{\volt}, \SI{-6.0}{\volt}]$ & \SI{-4.0}{\volt} & \SI{-4.5}{\volt}  \\
    Source current & $[\SI{11.8}{\milli \ampere}, \SI{1.0}{\milli \ampere}]$ & $\SI{6.4}{\milli \ampere}$ & $\SI{8.0}{\milli \ampere}$  \\
    Clear off &  $[\SI{4.5}{\volt}, \SI{8.0}{\volt}]$ & \SI{6.0}{\volt} & \SI{6.5}{\volt}\\
    Clear gate off &  $[\SI{4.2}{\volt}, \SI{8.0}{\volt}]$ & \SI{5.0}{\volt} &  \SI{5.0}{\volt} \\
    Clear on & $[\SI{13}{\volt}, \SI{27}{\volt}]$ & \SI{23.0}{\volt} & \SI{23.5}{\volt} \\
    Outer substrate & $[\SI{-10}{\volt}, \SI{0}{\volt}]$ & \SI{0}{\volt} & \SI{0}{\volt}\\ 
    Back contact & $[\SI{-114}{\volt}, \SI{-80}{\volt}]$ & \SI{-90.0}{\volt} & \SI{-90.0}{\volt}\\
    \hline
    \end{tabular}
    \label{tab:power-up sequence}
\end{table}

\FloatBarrier
\subsection{Variation of the back contact inner guard ring voltage}
\label{sec:BCIGR}
\FloatBarrier


The \textit{back contact inner guard ring} surrounds the entire sensitive area on the photon entrance side and is set to $V_{\text{BC\_IGR}} = \SI{-90}{\volt}$ by default, which is the same as the \textit{back contact} voltage.
It has been investigated whether a higher voltage has an effect on the amount of events registered on the outer edges of the DEPFET sensor. For this, $V_{\text{BC\_IGR}}$ is varied in the interval of $[\SI{-90}{\volt}, \, \SI{-100}{\volt}]$ in steps of \SI{2}{\volt}, with additional measurements at \SI{-105}{\volt} and \SI{-109}{\volt}.

\begin{figure}[htb!]
    \centering
    \includegraphics[width=\textwidth]{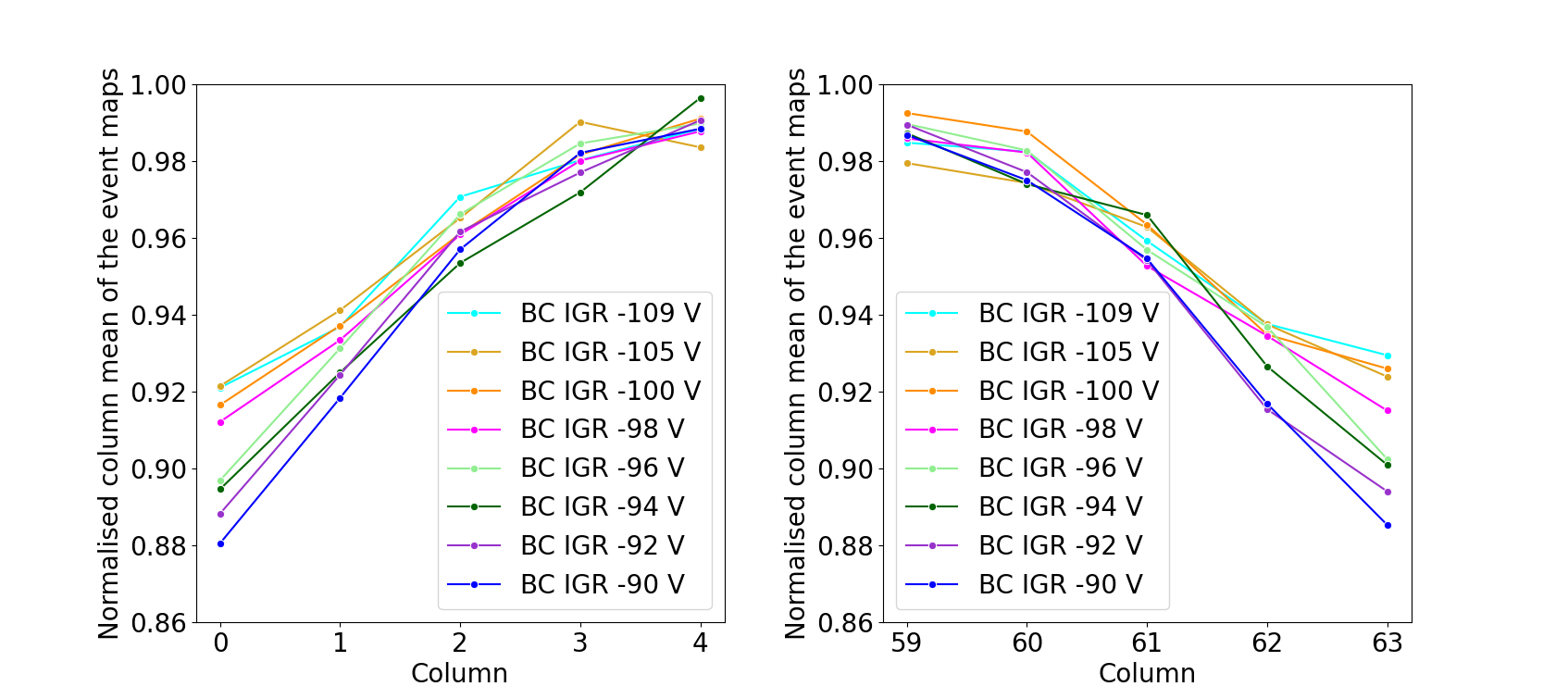}
    \caption{Column means of outermost left and right columns for the normalised event maps for different values of $V_{\text{BC\_IGR}}$ in the interval of $[\SI{-90}{\volt}, \, \SI{-100}{\volt}]$, as well as \SI{-105}{\volt} and \SI{-109}{\volt} (plotted as lines for better visibility).}
    \label{fig:BC_IGR_EventMaps}
\end{figure}

The round $^{55}$Fe source is positioned closely in front of the sensor but slightly off-centre during measurements, therefore inhomogeneities can be observed in the event map. The event map shows the number of photons detected for each pixel in the range of $[1 \, \mathrm{keV}, 14 \, \mathrm{keV}]$.
These inhomogeneities were compensated by performing a linear fit through the event map column mean values disregarding the outermost 8 columns and rows on both sides as here edge effects dominate over the inhomogeneities. 
The entire event map was then normalised using the fit data. Lastly, the column means of the outermost left and right columns were calculated and are shown in figure \ref{fig:BC_IGR_EventMaps} for all measurements.



For higher values of $V_{\text{BC\_IGR}}$ the column mean of the outermost columns is slightly higher compared to lower values of $V_{\text{BC\_IGR}}$, with the maximum difference of 0.044 between the measurements at \SI{-109}{\volt} and \SI{-90}{\volt}. Therefore, this effect is relatively small. It is possible to adapt $V_{\text{BC\_IGR}}$ to a value that differs from $V_{\text{BC}}$ in future versions of the \ac{PCM} in the electronics unit. 
For further measurements the default value for $V_{\text{BC\_IGR}}$ was left at \SI{-90}{\volt}.


\FloatBarrier
\label{sec:SRCDRN}
\FloatBarrier

\subsection{Variation of the source voltage}

Nominally the \textit{source} of the \ac{DEPFET} is set to a voltage of \SI{5.0}{\volt}, with a \textit{drain} voltage of \SI{1}{\volt} and \textit{gate} voltage of $\sim \SI{3}{\volt}$, so that a total current of \SI{6.4}{\milli \ampere} (which corresponds to \SI{100}{\micro \ampere} per pixel) runs through the DEPFETs of a selected row. 
For this measurement, the \textit{source} voltage $V_{\text{SRC}}$ was varied in the interval of $[\SI{4.0}{\volt}, \SI{7.0}{\volt}]$ in \SI{0.5}{\volt} steps, with additional measurements in smaller steps performed around the noise minimum. 
Other voltages powering parts of the DEPFET pixel, including the \textit{gate}, \textit{clear}, \textit{clear gate}, the \textit{rings}, and \textit{inner substrate}, were adjusted as they are defined with the \textit{source} as reference point. 
$V_{\text{SRC}}$ was then converted to the \textit{drain-source} voltage by calculating $V_{\text{DS}} = V_{\text{D}}-V_{\text{SRC}}$ with $V_{\text{D}} = \SI{1}{\volt}$. 
At lower temperatures the sensor is more prone to avalanche effects when large voltage differences are applied which cause higher dark currents.
Therefore, these measurements were repeated at two different sensor temperatures, $T \sim \SI{-60}{\celsius}$ and $T \sim \SI{-80}{\celsius}$.

\begin{figure}[htb!]
    \centering
    \includegraphics[width=0.6\linewidth]{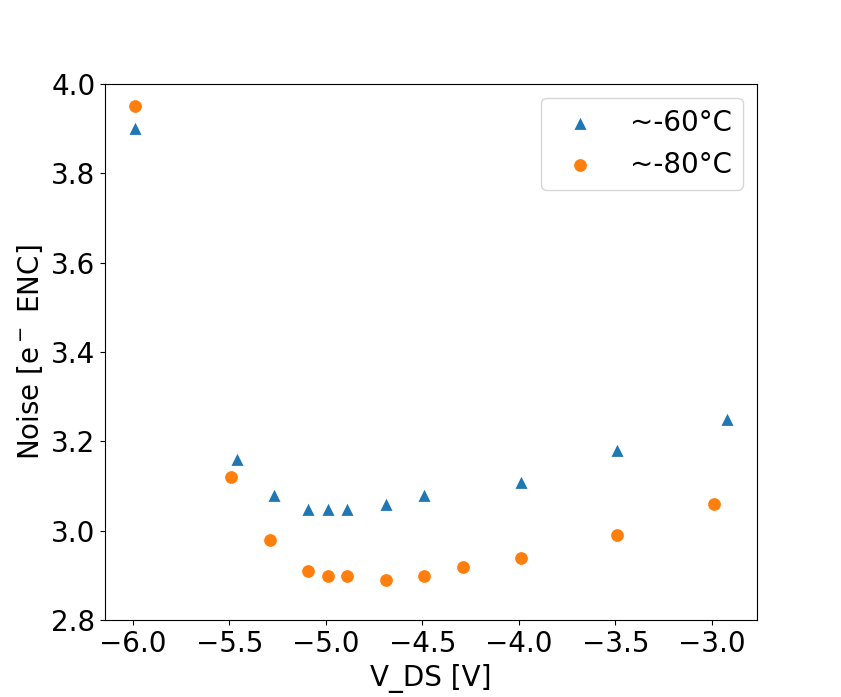}
    \caption{Noise [e$^-$ENC] of all measurements for variations of $V_{\text{DS}}$ at $\sim \SI{-60}{\celsius}$ and $\sim \SI{-80}{\celsius}$.}
    \label{fig:srcdrnTemp}
\end{figure}

Figure \ref{fig:srcdrnTemp} shows the noise [e$^-$ENC] of all measurements. It can be seen that the noise has a minimum for values of $V_{\text{DS}}$ between \SI{-4.5}{\volt} and \SI{-5.5}{\volt}. Additionally, the measurements at $\sim \SI{-80}{\celsius}$ result in lower noise values because of the lower dark current.

For further measurements $V_{\text{SRC}}$ was set to \SI{5.5}{\volt} which corresponds to $V_{\text{DS}} = \SI{-4.5}{\volt}$. This is slightly above the noise minimum, however the noise minimum is very close to the voltage where the noise starts to rise strongly. Setting the voltage slightly above ensures that even with fluctuations the sensor's performance with respect to the noise will stay excellent.

\subsection{Variation of the source current}

The \textit{source} current was indirectly probed by varying the \textit{gate on} voltage $V_{\text{Gate On}}$. For proper DEPFET operation the \textit{gate off} voltage and $V_{\text{VSSS}}$ have to be adjusted as well. $V_{\text{Gate On}}$ was varied in the interval of $ [\SI{2.62}{\volt}, \SI{4.62}{\volt}]$ in steps of $\SI{0.2}{\volt}$, resulting in \textit{drain-source} currents in the interval of $[\SI{11.8}{\milli \ampere}, \SI{1.0}{\milli \ampere}]$.

\begin{figure}[htb!]
    \centering
    \includegraphics[width=0.6\linewidth]{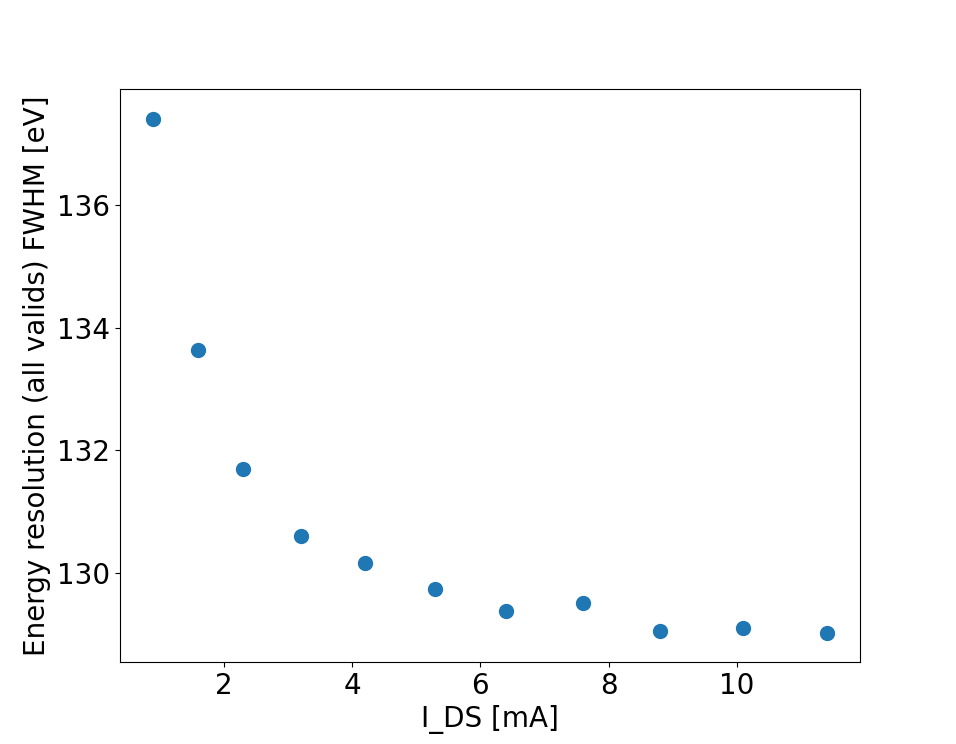}
    \caption{Energy resolution (FWHM of the Mn-K$\alpha$ peak) for the variation of $I_{\text{DS}}$.}
    \label{fig:A_Lo}
\end{figure}

Figure \ref{fig:A_Lo} shows the energy resolution (\ac{FWHM} of the Mn-K$\alpha$ peak) of all valid recombined events (so up to $2 \times 2$ pixels hit, and for multiple pixels the noise will increase compared to all valid single events). It can be seen that there are only minor improvements in noise and energy resolution once the \textit{drain-source} current is greater than about \SI{6}{\milli \ampere}. For lower values, an increase in \textit{source} current results in the degradation of energy resolution.

\FloatBarrier
\subsection{Variation of the clear off and clear gate off voltages}
\label{sec:BLoCLo}
\FloatBarrier

During these measurements, the influence of the \textit{clear off} and \textit{clear gate off} voltages on detector performance were investigated, especially with regards to the operational range and the optimal settings. 

The operational range is limited by two effects: First, \textit{back injection} occurs, when the \textit{clear gate} voltage is too high, compared to the \textit{clear} voltage. This results in charges flowing from the \textit{clear} contact into the \textit{internal gate}. Secondly, in this case, \textit{inversion} refers to when a conductive connection of holes under the \textit{clear gate} is created. As the \textit{clear gate} is an NMOS 
structure, electrons should be the primary charge carrier in the conductive channel. This conductive connection between the \ac{DEPFET} \textit{source} and \textit{drain} is caused by too low voltages applied to the \textit{clear gate}.

\begin{figure}[htb!]
    \centering
    \includegraphics[width=0.65\linewidth]{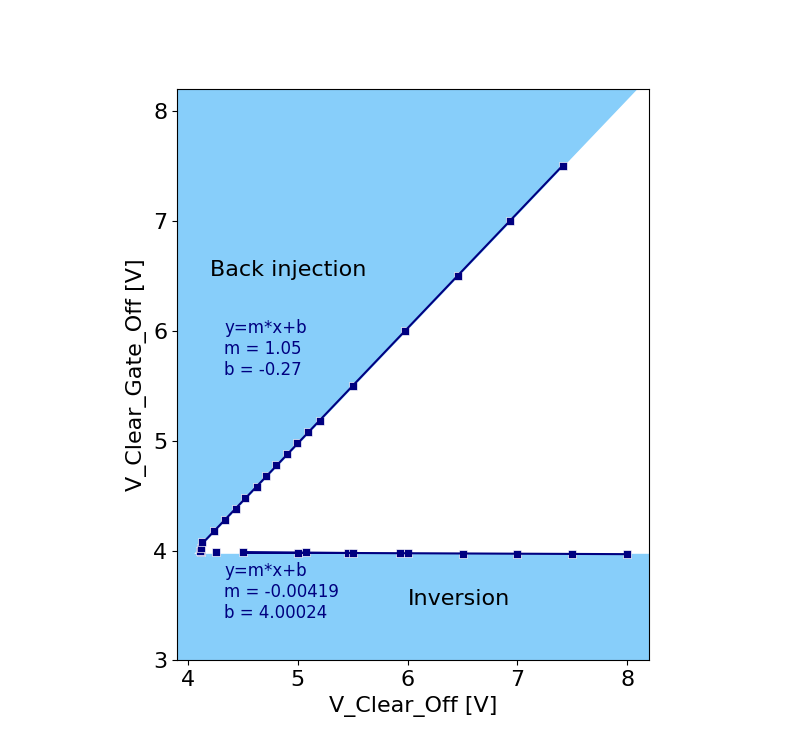}
    \caption{Fits through measurements (dark blue squares), limiting the operational window (white area).}
    \label{fig:BLoCLoLimits}
\end{figure}

To determine the limit of operation, different values of the \textit{clear} and \textit{clear gate} voltages were manually probed. When the voltage value is outside of the working range, the current flowing through the \ac{DEPFET} increases significantly which can be observed very clearly in the signal output from the VERITAS on an oscilloscope.
Values just before sensor operation became impossible were plotted and linear fits were performed to estimate the thresholds to the \textit{back injection} and \textit{inversion}. The result is shown in figure \ref{fig:BLoCLoLimits}.

Following this, a combined variation of the \textit{clear off} and \textit{clear gate off} voltages was performed. The \textit{clear off} voltage was varied in the interval of $[\SI{4.5}{\volt}, \SI{8.0}{\volt}]$ in steps of $\SI{0.5}{\volt}$, however only values resulting in an operational sensor were considered (essentially only \textit{clear off} voltages greater than the \textit{clear gate off} voltage were chosen). This variation was performed for the following \textit{clear gate off} voltages: $\SIlist{4.2; 5.0; 5.8; 6.6; 7.2; 8.0}{\volt}$. 

\begin{figure}[htb!]
    \centering
    \includegraphics[width=0.65\linewidth]{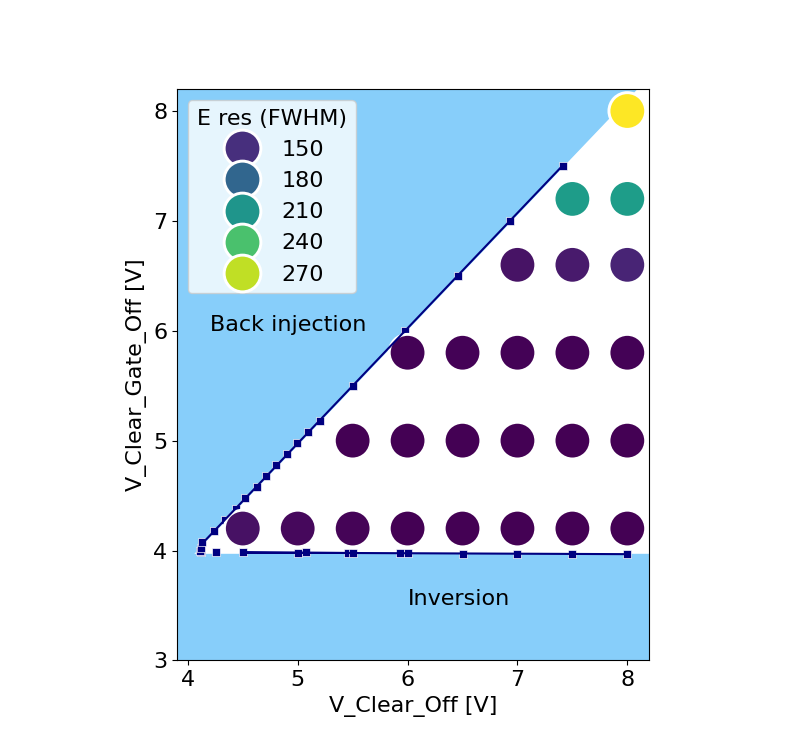}
    \caption{Energy resolution FWHM for variation of $V_{\text{Clear Gate Off}}$ and $V_{\text{Clear Off}}$.}
    \label{fig:BLoCLoEnergyRes}
\end{figure}

Figure \ref{fig:BLoCLoEnergyRes} shows the energy resolution \ac{FWHM} obtained from the different measurements. Overall, the energy resolution does not change significantly, especially for values sufficiently in the operational area of the sensor. 
The sensor performance decreases for very high values of $V_{\text{Clear Gate Off}}$. It is suspected, that the uppermost point is affected by the point being outside of the operational area and instead in the area of \textit{back injection}, or that there is clearing even though the \textit{clear} is supposed to be off. 
It is also possible that charge carriers start draining to the \textit{clear} which therefore appears as apparent gain reduction. 
Both the default values of $V_{\text{Clear Gate Off}} = \SI{5.0}{\volt}$ and $V_{\text{Clear Off}} =  \SI{6.0}{\volt} \text{ or }  \SI{6.5}{\volt}$ have an excellent energy resolution.




\FloatBarrier
\subsection{Variation of the clear on voltage}
\label{sec:BHi}
\FloatBarrier

In the past, effects have been observed that are thought to stem from the Switcher-A \ac{ASIC} emitting light for large differences between the \textit{clear on} and \textit{clear off} voltages. Therefore a variation of the \textit{clear on} voltage is performed with constant \textit{clear off} voltage to investigate the behaviour and find the limit as to where the performance would be impacted.

For this, a variation was performed in two steps, first with measurements in the interval of $[\SI{13}{\volt}, \SI{27}{\volt}]$ in steps of $\SI{2}{\volt}$ and then $[\SI{23}{\volt}, \SI{25}{\volt}]$ in steps of $\SI{0.2}{\volt}$ to find the point where the performance decreases. The \textit{clear gate off} voltage was set to $\SI{6.5}{\volt}$ for all measurements.

\begin{figure}[H]
    \centering
    \includegraphics[width=\linewidth]{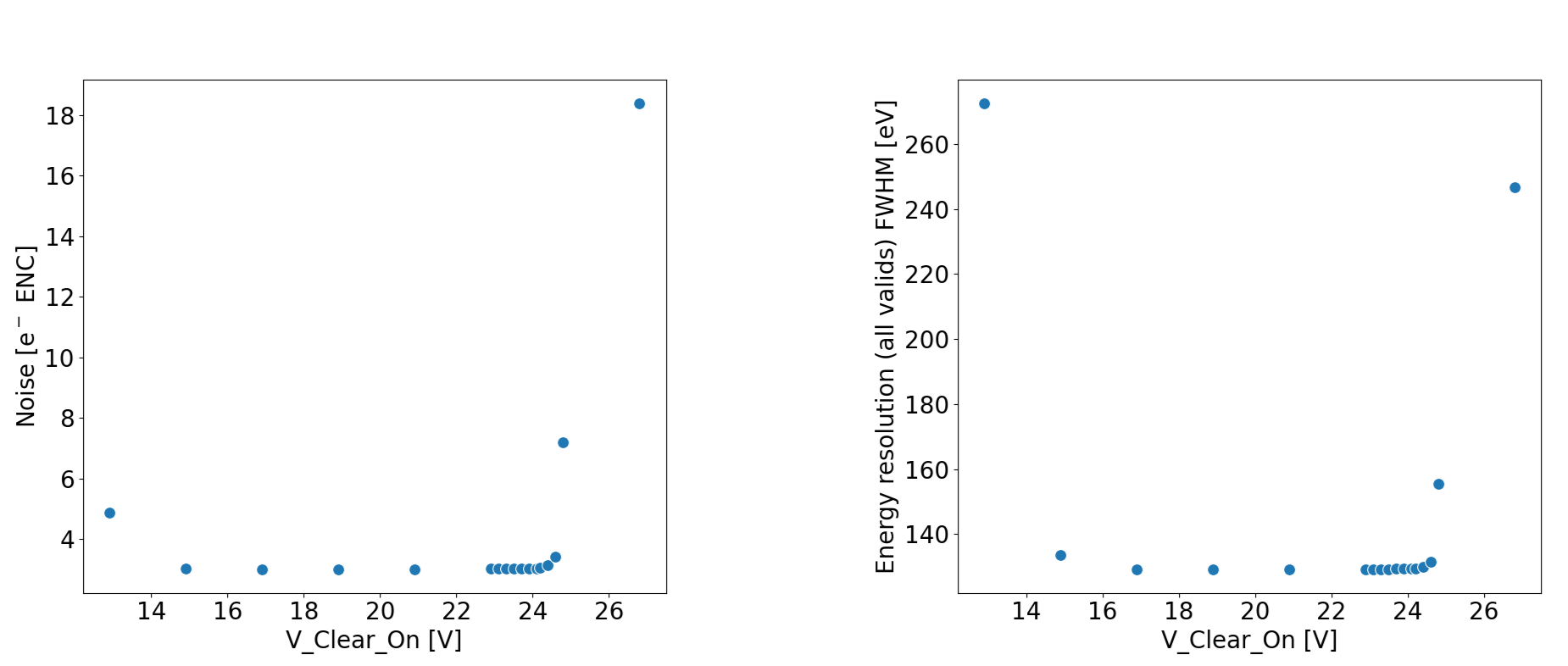}
    \caption{Noise and energy resolution (FWHM of the Mn-K$\alpha$ peak) for the variation of the \textit{clear on} voltage.}
    \label{fig:BHiPlots}
\end{figure}

Figure \ref{fig:BHiPlots} shows the noise [e$^-$ENC], gain and energy resolution for this variation. The sensor performance starts to decrease above $\SI{24}{\volt}$, as can be seen by the sharp increase in noise and energy resolution. Therefore the \textit{clear gate on} voltage should be set to a value $\le \SI{24}{\volt}$ to ensure good sensor performance.

To verify that the decrease in sensor performance is caused by a part of the Switcher-A \ac{ASIC} starting to glow as suspected, a series of measurements was performed at $V_{\text{Clear On}}=\SI{25}{\volt}$ for different exposure times. The exposure times chosen were multiples of  \SI{1267.5}{\micro \second}. First $\{1, 2, 3, 4, 5, 6, 7, 8\} \cdot \SI{1267.5}{\micro \second}$ and then additional measurements at $\{10, 20, 30, 40, 50\} \cdot \SI{1267.5}{\micro \second}$.

\begin{figure}[htb!]
    \centering
    \includegraphics[width=0.6\linewidth]{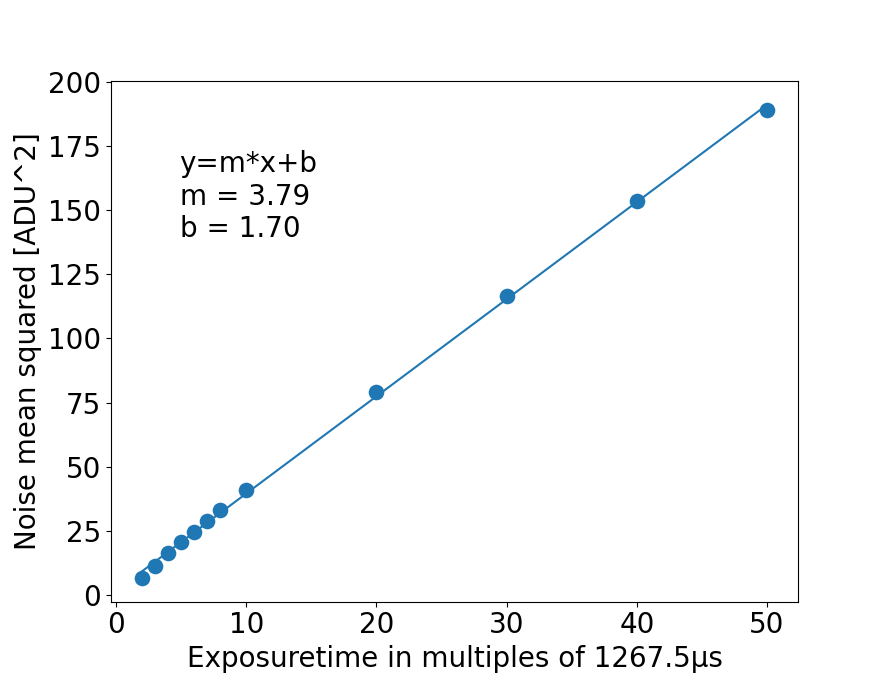}
    \caption{Noise mean squared in $[\text{ADU}^2]$ for different exposure times (points) and linear fit through these points (line).}
    \label{fig:BHi_Exposure}
\end{figure}

The result of the measurement is shown in figure \ref{fig:BHi_Exposure} which shows the noise mean squared, which is proportional to the 
noise mean squared for the different exposure times expressed as multiples of \SI{1267.5}{\micro \second}. A linear fit was performed, resulting in the following function: $y=3.79 \cdot x + 1.70$. The measurements showing good linear behaviour, except for very short exposure times.

This linear behaviour is consistent with the assumption that the decrease in sensor performance is caused by the Switcher-A starting to glow at high voltages. 
This is because charge carriers have to be continuously created for this linear behaviour to occur. As the parameters affecting the sensors, especially the temperature, were not changed, the linear behaviour is unlikely to be caused by other effects increasing the dark current. Additionally, higher noise can also be seen in the area close to the Switcher-A in the calibrated noise map, which is also consistent with light emissions from the Switcher-A.
As a new Switcher-A \ac{ASIC} version is in development, this test has to be repeated in the future. For the \ac{LD} and \ac{FD} sensors for use in the Camera Head this matters less as their \acp{ASIC} will be installed on the same height as the sensor and therefore the light will mainly shine onto the non sensitive area of the sensor. On the prototype modules, the \acp{ASIC} are installed higher, relative to the sensor, which makes it easier for light to reach the sensitive area of the sensor.

For further measurements the voltage was set to $V_{\text{Clear On}}=\SI{23.5}{\volt}$.

\FloatBarrier
\subsection{Variation of the outer substrate voltage}
\label{sec:OS}
\FloatBarrier

As default, the \textit{\acf{OS}} is set to $\SI{0}{\volt}$. A variation of this voltage was performed to see if an improvement of performance in the outermost pixels could be observed, similar motivation to why an investigation of the \textit{back contact inner guard ring} was performed in section \ref{sec:BCIGR}. 
For this the voltage was varied in an interval of $[\SI{-10}{\volt}, \SI{0}{\volt}]$ in steps of $\SI{1}{\volt}$. 

\begin{figure}[htb!]
    \centering
    \includegraphics[width=\linewidth]{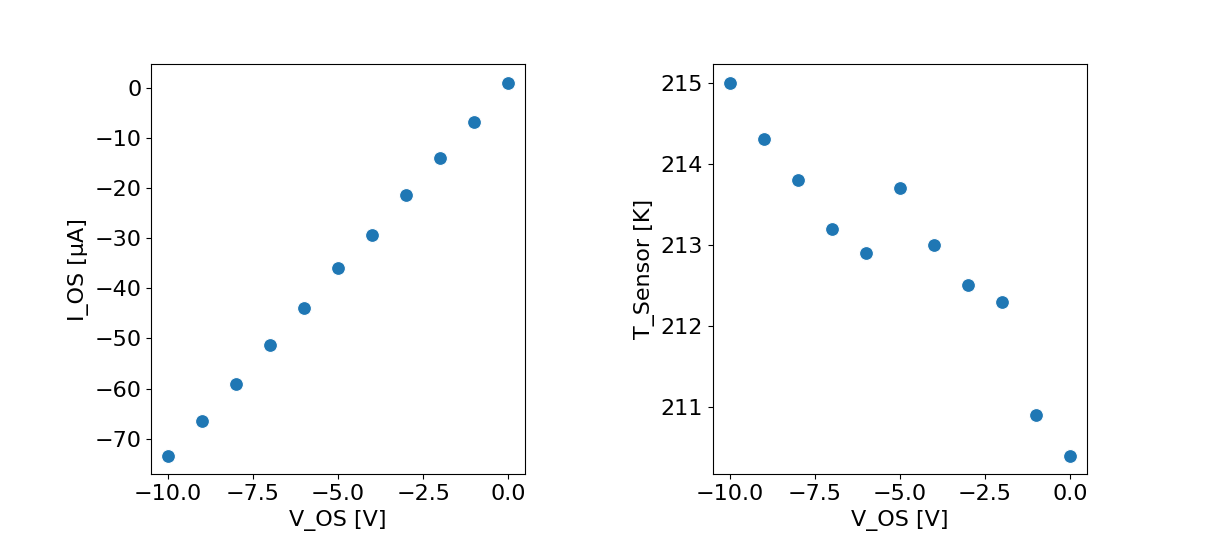}
    \caption{\textit{Left:} \textit{OS} current for different values of \textit{OS} voltage, \textit{right:} sensor temperature for different values of \textit{OS} voltage.}
    \label{fig:OS}
\end{figure}

A setting for improved performance in the outermost pixels could not be found and instead an increase in current was measured for lower values of the \textit{OS} voltage. This can be seen in the left plot of figure \ref{fig:OS}. This is likely caused by the temperature diode being located close to the \textit{OS} and the current from the \textit{OS} likely affecting the diode and therefore the temperature value extracted from measurement. The steps in the temperature measurement for the different \textit{OS} voltages shown in the right plot in figure \ref{fig:OS} were probably caused by the measurements not being performed in one go but at different times during the afternoon.

\FloatBarrier
\subsection{Variation of the back contact voltage}
\label{sec:BC}
\FloatBarrier

The \textit{back contact} voltage is typically set to $V_{\text{BC}} = \SI{-90}{\volt}$, same as the \textit{back contact inner guard ring} varied in section \ref{sec:BCIGR}. For this measurement, the \textit{back contact} voltage was varied from $\SI{-80}{\volt}$ to $\SI{-114}{\volt}$ in steps of $\SI{5}{\volt}$ (where possible). The voltage of the \textit{back contact inner guard ring} was set to the same value as the voltage set to the \textit{back contact}.

\begin{figure}[htb!]
    \centering
    \includegraphics[width=0.8\linewidth]{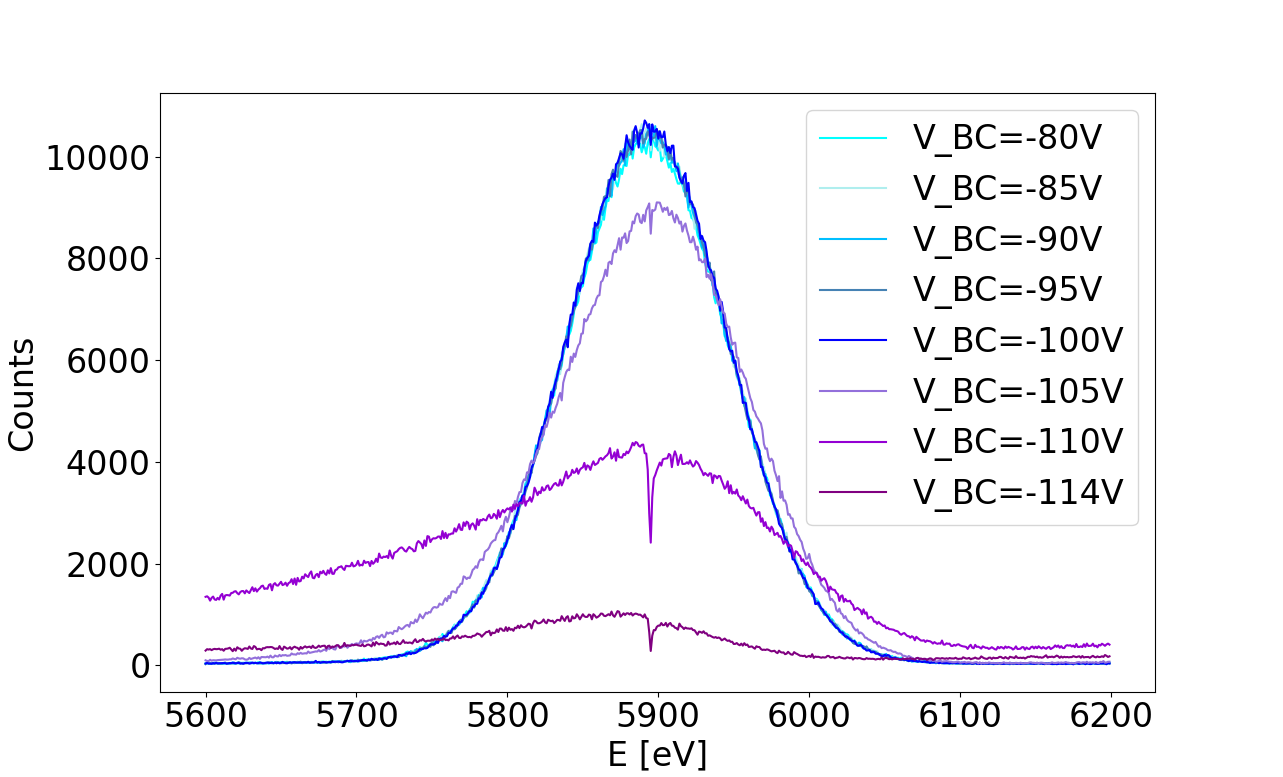}
    \caption{Energy spectrum around the Mn-K$\alpha$ peak for different values of $V_{\text{BC}}$.}
    \label{fig:BCKalpha}
\end{figure}

Figure \ref{fig:BCKalpha} shows the energy spectrum around the Mn-K$\alpha$ peak for the different values of $V_{\text{BC}}$. It can be seen that there is no big difference in performance for the more positive values of $V_{\text{BC}}$. For more negative values the energy resolution increases and the Mn-K$\alpha$ peak is a lot wider and asymmetric with more counts registered for lower energies. This is due to the loss of charge carriers to the \textit{clear} and therefore not being measured by the pixel. 
From these measurements it seems that there is no significant difference in performance for values of $V_{\text{BC}} \in [\SI{-80}{\volt}, \SI{-100}{\volt}]$, therefore for further measurements $V_{\text{BC}}$ was left at $\SI{-90}{\volt}$.

\FloatBarrier
\section{Timing Variation}
\label{sec:SRCDRN}
\FloatBarrier

The duration of the \textit{settling} and \textit{clear} stages was varied to get insights into the areas of best performance and limits. For this, \textit{settling 1}, \textit{settling 2}, and the \textit{clear} duration were individually varied and finally \textit{settling 2} and \textit{clear} were varied together in a way that they both were set to the same time. All other variables were set to \SI{1000}{\nano \second} unless stated otherwise. 
The time between the \textit{clear gate} and the \textit{clear} being turned off was always set to \SI{25}{\nano \second}. This is to ensure that charges in the \textit{clear gate} at the time that the \textit{clear gate} is shut off propagate to the \textit{clear} and not to the \textit{internal gate}.

For each measurement both dark frames and photon frames were taken and analysed. Offset values (minimum, maximum, and mean) are calculated with respect to all $64 \times 64$ pixels.

For sensor operation the goal is to have a small difference between the offset minimum and maximum. This is because the dynamic range should be maximised. The 14 bit \ac{ADC} has to fit both the difference between offset maximum and minimum plus noise including the common mode noise and the dynamical range. A small value for the span from offset maximum to minimum therefore results in much better binning of the entire energy range.

\subsection{Variation of settling 1 duration}
\label{subsec:sett1}
\FloatBarrier
The first \textit{settling} was varied in the interval of $[\SI{200}{\nano \second}, \SI{1000}{\nano \second}]$ in steps of \SI{100}{\nano \second}.

\begin{figure}[H]
    \centering
    \includegraphics[width=0.6\linewidth]{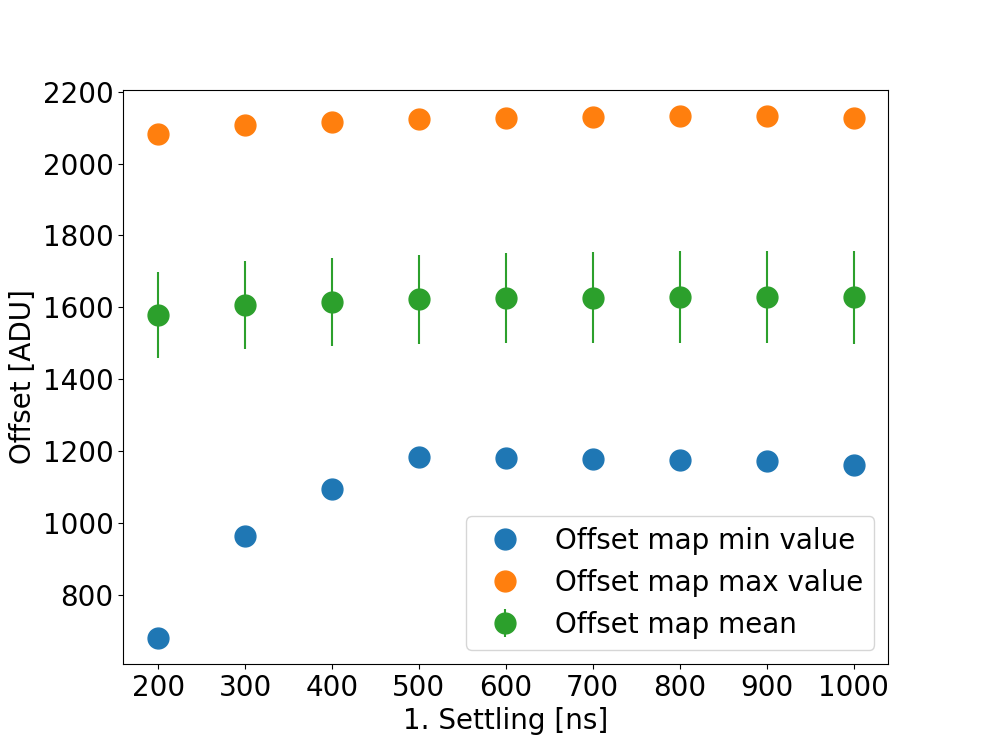}
    \caption{Mean, minimum, and maximum value of the offset map for different \textit{settling 1} durations.}
    \label{fig:OffsetSett1}
\end{figure}

Figure \ref{fig:OffsetSett1} shows the mean, minimum, and maximum offset values for the different measurements. Both the maximum value as well as the mean stay relatively constant for all \textit{settling 1} durations. The minimum value stays almost constant for values of and above \SI{500}{\nano \second} and decreases for lower values.
This suggests that for optimal performance, the duration of \textit{settling 1} should be set to at least \SI{500}{\nano \second}. Additionally, no significant improvements were observed in the noise, gain, or energy resolution for longer \textit{settling 1} times than \SI{700}{\nano \second}. 

\FloatBarrier
\subsection{Variation of settling 2 duration}
\label{subsec:sett2}
\FloatBarrier
The second \textit{settling} was varied in the interval of $[\SI{100}{\nano \second}, \SI{1000}{\nano \second}]$ in steps of \SI{100}{\nano \second}.

\begin{figure}[H]
    \centering
    \includegraphics[width=0.6\linewidth]{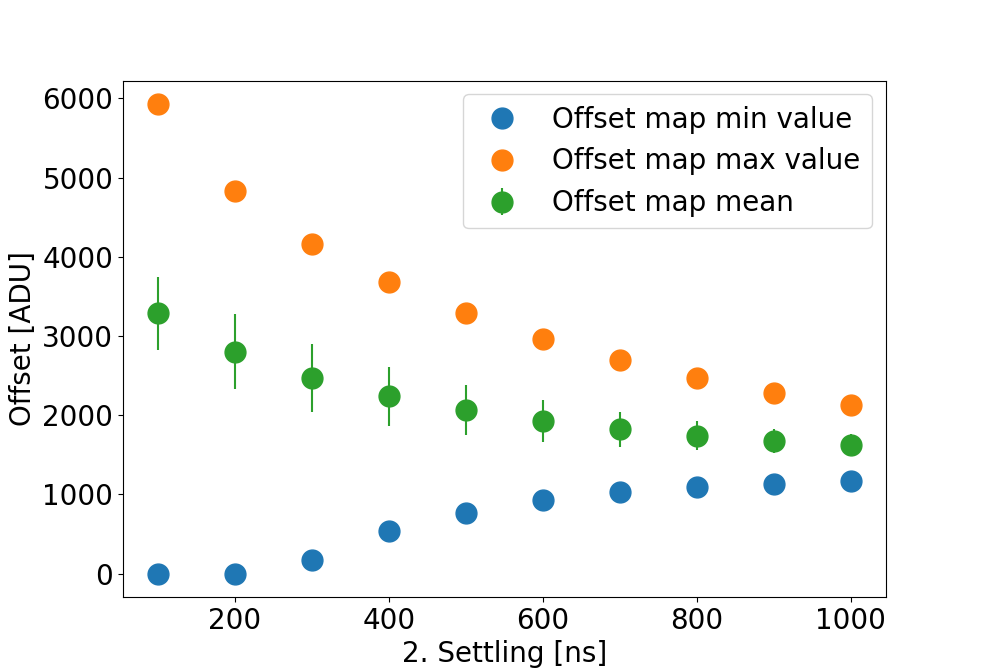}
    \caption{Mean, minimum, and maximum value of the offset map for different \textit{settling 2} durations.}
    \label{fig:OffsetSett2}
\end{figure}

Figure \ref{fig:OffsetSett2} shows the mean, minimum, and maximum offset values for the different measurements. Both the maximum as well as the mean value decrease with longer \textit{settling 2} duration. The minimum value increases but less strongly than the maximum value decreases. The standard deviation also decreases for longer durations. The offset mean and maximum are also significantly larger compared to the \textit{settling 1} variation.

\FloatBarrier
\subsection{Variation of the clear duration}
\label{subsec:clear}
\FloatBarrier

For the variation of the \textit{clear} duration the \textit{clear} correlation amplitude was evaluated. The \textit{clear} correlation amplitude is higher if there is an incomplete \textit{clear}, so if there are charge carriers from a previous frame those will show up in the following frame and result in a higher \textit{clear} correlation amplitude. Therefore a low value for the \textit{clear} correlation amplitude is desired as this means few charge carriers appear in the pixel in the following frames. 

\begin{figure}[H]
    \centering
    \includegraphics[width=0.6\linewidth]{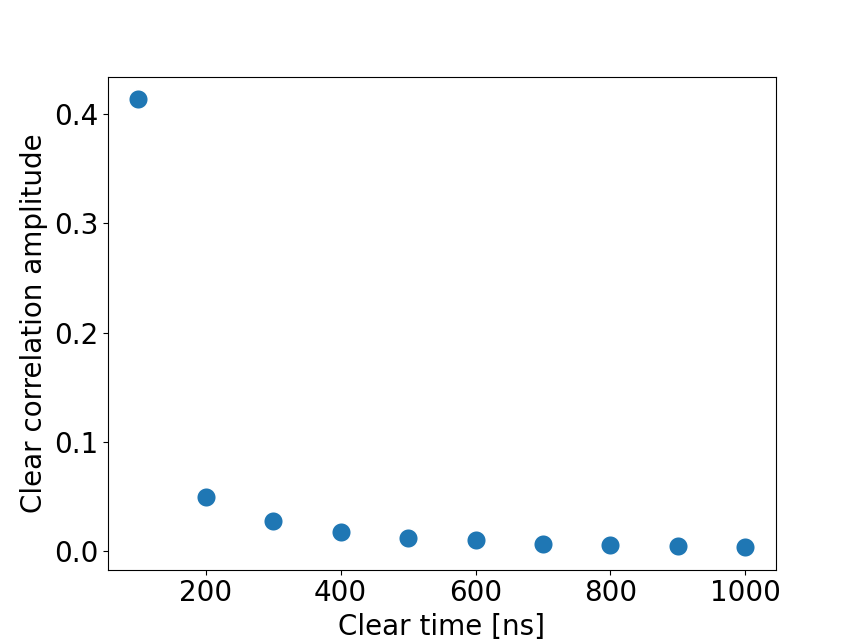}
    \caption{\textit{Clear} correlation amplitude for different \textit{clear} durations.}
    \label{fig:OffsetClear}
\end{figure}

The \textit{clear} duration was varied in the interval of $[\SI{100}{\nano \second}, \SI{1000}{\nano \second}]$ in steps of \SI{100}{\nano \second}.
Figure \ref{fig:OffsetClear} shows the \textit{clear} correlation amplitude for the different \textit{clear} timings. For longer \textit{clear} timings the \textit{clear} correlation amplitude goes down, with the gradient being steeper for lower \textit{clear} times. For values of $\SI{700}{\nano \second}$ or higher the \textit{clear} correlation amplitude settles. 
For measurements using the \ac{LD}, a longer \textit{clear} time is preferred, especially after irradiation or use in space as it is expected that the \textit{clear} degrades. 


\FloatBarrier
\subsection{Variation of the settling 2 and clear duration simultaneously}

Lastly, \textit{settling 2} and the \textit{clear} time were varied simultaneously in the interval of $[\SI{100}{\nano \second}, \SI{1000}{\nano \second}]$ in steps of \SI{100}{\nano \second} in a way that they were always set to the same value.  

\begin{figure}[H]
    \centering
    \includegraphics[width=0.6\linewidth]{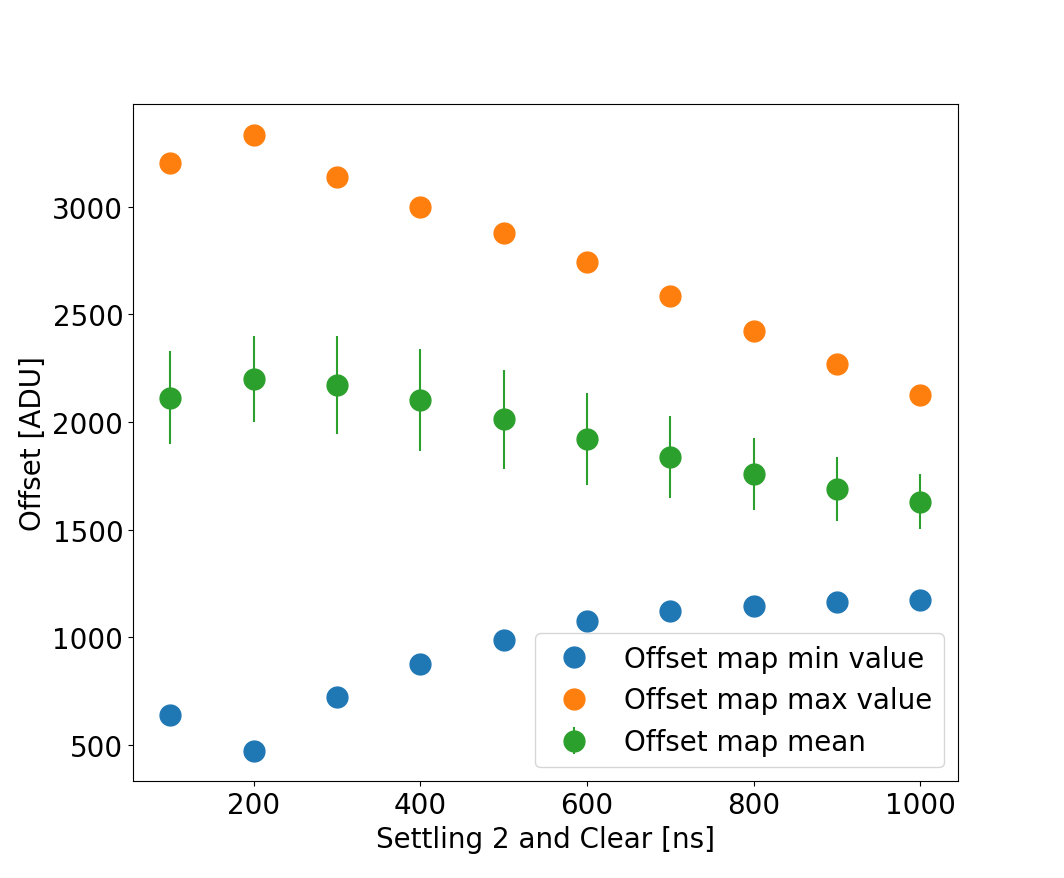}
    \caption{Mean, minimum, and maximum value of the offset map for different \textit{settling 2} and \textit{clear} durations.}
    \label{fig:OffsetSett2Clear}
\end{figure}

Figure \ref{fig:OffsetSett2Clear} shows the mean, minimum, and maximum value of the offset map for there different timing settings. The difference between the offset map minimum and maximum decreases for longer timings, similar to what was observed if only \textit{settling 2} was varied in section \ref{subsec:sett2}. Even for fast timings, the difference between offset maximum and minimum is smaller than if only \textit{settling 2} is varied.

\FloatBarrier

\subsection{Timing conclusions}

From the parameters that were studied it seems the best choice for optimised performance is to set \textit{settling 1}, \textit{2}, and the \textit{clear} to \SI{1000}{\nano \second}, or at least until they are mostly settled. However, other effects also contribute, some of which have been studied in \ccite{Treberspurg_2018, Treberspurg_2019}. For example longer frame times lead to larger noise because more dark current can accumulate. Additionally, the time between the first and second integration should be kept as short as possible to reduce the $1/f$-noise, therefore it is better to choose values where \textit{settling 1}, \textit{2}, and the \textit{clear} are more or less settled but not longer.
Lastly, for the \ac{FD} the requirements to achieve the frame time of \SI{80}{\micro \second} don't allow for the most optimised timing values to be chosen.

\section{Conclusions}
\label{sec:conc}
\FloatBarrier

The main voltages and currents affecting the DEPFET sensor were varied and analysed for optimised performance and limits of operation. 

The most important result is that none of the parameters varied showed small working areas for the detector, but rather that good performance was possible over wide ranges. 
This means that there is margin for pixel-wise variations over the many pixels of the \ac{LDA} even if their operational parameters spread further due to an inhomogeneous degradation after long use in the space environment, as seen in irradiation studies \ccite{Valentin}. While slight global adaptions can compensate a global shift in voltages, there should be no limit in the operation voltages that decreases the spectral performance significantly.

Small improvements in the performance of the outermost columns were observed when the voltage set to the \textit{back contact inner guard ring} was increased to more negative values than the nominal value of $\SI{-90}{\volt}$.
The noise could be reduced by increasing the \textit{source} voltage from $V_{\text{SRC}}= \SI{5.0}{\volt}$ to $V_{\text{SRC}}= \SI{5.5}{\volt}$ which corresponds to a \textit{drain-source} voltage of $V_{\text{DS}} = \SI{-4.5}{\volt}$.
Only minor improvements in noise and energy resolution could be observed once the \textit{source} current is set to a value greater than about \SI{6}{\milli \ampere}. The \textit{source} current was set to $I_{\text{DS}}=\SI{8.0}{\milli \ampere}$.
The \textit{clear off} and \textit{clear gate off} voltages were investigated and the thresholds to back injection and inversion were identified. Inside the operational area the difference in performance was mostly insignificant, for ideal performance the value for the \textit{clear off} should be increased by $\SI{0.5}{\volt}$ to $V_{\text{Clear Off}} =  \SI{6.5}{\volt}$. 
The \textit{clear on} voltage was also varied and for high voltages dark current increases due to light emissions from the ASIC were observed. The decrease in performance can be avoided for values below $V_{\text{Clear On}} =  \SI{24}{\volt}$. For further measurements the value was set to $V_{\text{Clear On}} =  \SI{23.5}{\volt}$.
The voltage applied to the \textit{outer substrate} was varied, however applying more negative voltages led to currents flowing to the temperature diode.
For very negative values of the voltage applied to the \textit{back contact}, so for $V_{\text{BC}} < \SI{-100}{\volt}$, the energy resolution decreases due to the loss of charge carriers towards the \textit{clear}. The value was therefore left at $V_{\text{BC}} = \SI{-90}{\volt}$.
Different timing settings were varied. From these measurements, it seems that longer times for \textit{settling 1}, \textit{settling 2}, and the \textit{clear} increase performance, however other parameters, e.g. the timing requirements for the FD, also have to be taken into account when choosing timing values.

To study the impact of the new sensor settings measurements were conducted with the old and new default values. The most important results are shown in table \ref{tab:vglaltneu}. In both noise and energy resolution slight improvements can be observed. For the energy resolution the difference is negligible.

\begin{table}[htb]
    \centering
    \caption{Noise and energy resolution results of measurements with old and new default settings.}
    \begin{tabular}{|c|c|c|}
    \hline
     & Old default settings &  New default settings\\
     \hline
     Noise [e$^-$ENC] & $2.973 \pm 0.003$  & $ 2.905 \pm 0.003$ \\
     Energy resolution all (FWHM) & $129.5 \pm 0.1$ & $ 128.0 \pm 0.1 $ \\
    \hline
    \end{tabular}
    \label{tab:vglaltneu}
\end{table}

This means that the parameters of the pre-flight production are well reproduced by the flight production of the DEPFETs. Therefore, apart from slight optimisations, no adjustments are needed. This reproducibility highlights the quality of the sensor productions of the MPG semiconductor laboratory. Additionally, limits to the operational range were identified, and also a closer understanding was gained of where the sensor still operates with good but not ideal performance. 
The measurements varying ASIC voltages, especially regarding the glowing Switcher-A conducted in \ref{sec:BHi} will have to be conducted again with future versions of the ASIC and relevant geometry changes.

\FloatBarrier

\appendix    

\acknowledgments 
We are grateful to the whole Athena WFI project team and all MPE colleagues, who contributed to the development, assembly and testing. 
Development and production of the DEPFET sensors for the Athena WFI is performed in a collaboration between MPE and the MPG Semiconductor Laboratory (HLL). The work was funded by the Max Planck Society and the German space agency DRL (FZK: 50 QR 2501).

\bibliography{report} 

@misc{nandra2013hotenergeticuniversewhite,
      title={The Hot and Energetic Universe: A White Paper presenting the science theme motivating the Athena+ mission}, 
      author={{Kirpal Nandra et al.}},
      year={2013},
      eprint={1306.2307},
      archivePrefix={arXiv},
      primaryClass={astro-ph.HE},
      url={https://arxiv.org/abs/1306.2307}, 
}

@inproceedings{Müller_Seidlitz_2024,
author = {J. M{\"u}ller-Seidlitz and R. Andritschke and V. Emberger and M. Bonholzer and G. Hauser and P. Lechner and A. Mayr and J. Reiffers and A. Schweingruber and W. Treberspurg},
title = {{Spectroscopic performance of detectors for Athena’s WFI: measurements and simulation}},
volume = {13093},
booktitle = {Space Telescopes and Instrumentation 2024: Ultraviolet to Gamma Ray},
editor = {Jan-Willem A. den Herder and Shouleh Nikzad and Kazuhiro Nakazawa},
organization = {International Society for Optics and Photonics},
publisher = {SPIE},
pages = {130930T},
year = {2024},
doi = {10.1117/12.3019707},
URL = {https://doi.org/10.1117/12.3019707}
}

@inproceedings{SPOpaper,
author = {{Maximilien J. Collon et al.}},
title = {{Silicon pore x-ray optics for the NewAthena telescope}},
volume = {13093},
booktitle = {Space Telescopes and Instrumentation 2024: Ultraviolet to Gamma Ray},
editor = {Jan-Willem A. den Herder and Shouleh Nikzad and Kazuhiro Nakazawa},
organization = {International Society for Optics and Photonics},
publisher = {SPIE},
pages = {1309319},
year = {2024},
doi = {10.1117/12.3019675},
URL = {https://doi.org/10.1117/12.3019675}
}

@misc{nandratobepublished,
      title={}, 
      author={{Kirpal Nandra et al.}},
      year={in preparation, 2027},
      eprint={},
      archivePrefix={},
      primaryClass={},
      url={}, 
}

@inproceedings{SpieValeria,
author = {Valeria Antonelli and Daniel Pietschner and Rafael Strecker and Benjamin Mican and Jan Philipp M{\"o}ller and Aysun S{\"o}nmez and Anirudh Saraf and Hermine Schnetler and Paul Nandra},
title = {{The Wide Field Imager for the NewAthena mission: preliminary design and verification}},
volume = {13093},
booktitle = {Space Telescopes and Instrumentation 2024: Ultraviolet to Gamma Ray},
editor = {Jan-Willem A. den Herder and Shouleh Nikzad and Kazuhiro Nakazawa},
organization = {International Society for Optics and Photonics},
publisher = {SPIE},
pages = {130934L},
year = {2024},
doi = {10.1117/12.3020214},
URL = {https://doi.org/10.1117/12.3020214}
}

@inproceedings{SpieAstrid,
author = {Astrid Mayr and Johannes M{\"u}ller-Seidlitz and Valentin Emberger and Robert Andritschke and Jonas Reiffers and Anna Schweingruber and Sebastian Albrecht and Hermine Schnetler},
title = {{Spectral performance budget for ATHENA’s Wide Field Imager}},
volume = {13099},
booktitle = {Modeling, Systems Engineering, and Project Management for Astronomy XI},
editor = {S{\'e}bastien Elias Egner and Scott Roberts},
organization = {International Society for Optics and Photonics},
publisher = {SPIE},
pages = {130992E},
year = {2024},
doi = {10.1117/12.3018737},
URL = {https://doi.org/10.1117/12.3018737}
}

@misc{WFIinternal,
  author = "{WFI consortium}",
  howpublished = "Private communication"
}

@article{Treberspurg_2018,
doi = {10.1088/1748-0221/13/12/P12001},
url = {https://doi.org/10.1088/1748-0221/13/12/P12001},
year = {2018},
month = {December},
publisher = {},
volume = {13},
number = {12},
pages = {P12001},
author = {Treberspurg, W. and Meidinger, N. and Müller-Seidlitz, J. and Herrmann, S.},
title = {Achievable noise performance of spectroscopic prototype DEPFET detectors},
journal = {Journal of Instrumentation}
}

@article{Treberspurg_2019,
doi = {10.1088/1748-0221/14/03/P03019},
url = {https://doi.org/10.1088/1748-0221/14/03/P03019},
year = {2019},
month = {March},
publisher = {},
volume = {14},
number = {03},
pages = {P03019},
author = {Treberspurg, W. and Hauser, G. and Meidinger, N. and Müller-Seidlitz, J. and Ott, S.},
title = {Achievable time resolution of spectroscopic prototype DEPFET detectors},
journal = {Journal of Instrumentation}
}

@inproceedings{Valentin,
author = {Valentin Emberger and Robert Andritschke and Parviz Azhdarzadeh and G{\"u}nter Hauser and Astrid Mayr and Johannes M{\"u}ller-Seidlitz and Abbas Rezaei and Wolfgang Treberer-Treberspurg},
title = {{Low-temperature proton irradiation with DEPFETs for Athena’s wide field imager}},
volume = {13093},
booktitle = {Space Telescopes and Instrumentation 2024: Ultraviolet to Gamma Ray},
editor = {Jan-Willem A. den Herder and Shouleh Nikzad and Kazuhiro Nakazawa},
organization = {International Society for Optics and Photonics},
publisher = {SPIE},
pages = {130930S},
year = {2024},
doi = {10.1117/12.3018769},
URL = {https://doi.org/10.1117/12.3018769}
}

@thesis{JohannesPhD,
          author = {Johannes M{\"u}ller-Seidlitz},
            year = {2020},
           title = {A megahertz active pixel sensor for X-ray astronomy},
           month = {January},
       publisher = {Ludwig-Maximilians-Universit{\"a}t M{\"u}nchen},
             url = {http://nbn-resolving.de/urn:nbn:de:bvb:19-254947}
}

@misc{ROAn,
      title={ROAn, a ROOT based Analysis Framework}, 
      author={Thomas Lauf and Robert Andritschke},
      year={2013},
      eprint={1310.4848},
      archivePrefix={arXiv},
      primaryClass={physics.data-an},
      url={https://arxiv.org/abs/1310.4848}, 
}

@inproceedings{Veritas1,
author = {Sven Herrmann and Anna Koch and Sara Obergassel and Wolfgang Treberspurg and Michael Bonholzer and Norbert Meidinger},
title = {{VERITAS 2.2: a low noise source follower and drain current readout integrated circuit for the wide field imager on the Athena x-ray satellite}},
volume = {10709},
booktitle = {High Energy, Optical, and Infrared Detectors for Astronomy VIII},
editor = {Andrew D. Holland and James Beletic},
organization = {International Society for Optics and Photonics},
publisher = {SPIE},
pages = {1070935},
year = {2018},
doi = {10.1117/12.2314335},
URL = {https://doi.org/10.1117/12.2314335}
}

@inproceedings{VeritasAnna,
author = {Anna-Katharina Schweingruber and Sven Herrmann and Peter Orel and Ajay Kumar Dakshinamurthy and Astrid Mayr and Johannes M{\"u}ller-Seidlitz and Jonas Reiffers and Sebastian Albrecht and Hermine Schnetler and Steven W. Allen and Glenn Morris},
title = {{The VERITAS 2.3 readout ASIC for the ATHENA Wide Field Imager}},
volume = {13093},
booktitle = {Space Telescopes and Instrumentation 2024: Ultraviolet to Gamma Ray},
editor = {Jan-Willem A. den Herder and Shouleh Nikzad and Kazuhiro Nakazawa},
organization = {International Society for Optics and Photonics},
publisher = {SPIE},
pages = {130934D},
year = {2024},
doi = {10.1117/12.3018879},
URL = {https://doi.org/10.1117/12.3018879}
}
\bibliographystyle{spiebib} 

\end{document}